\documentclass[a4paper,11pt]{article}
\pdfoutput=1
\usepackage{jheppub}

\usepackage{amsmath} 
\usepackage{graphicx} 
\usepackage{amsthm}
\usepackage{amssymb} 
\usepackage{dsfont}
\usepackage{yfonts}
\usepackage{hyperref}
\usepackage{floatrow,dcolumn,rotating}
\usepackage{array,xcolor,graphicx}
\usepackage{booktabs,multirow}
\usepackage[utf8]{inputenc}
\usepackage{mathtools}

\usepackage[normalem]{ulem}

\usepackage[all]{xy}
\usepackage{tikz}
\usetikzlibrary{arrows.meta}

\newcommand{\be}{\begin{equation}}
\newcommand{\ee}{\end{equation}}
\newcommand{\bea}{\begin{eqnarray}}
\newcommand{\eea}{\end{eqnarray}}

\newcommand{\nn}{\nonumber\\}
\def\vecb#1{\mathbf{#1}}
\def\CC{\mathcal{C}}

\def\CN{\mathcal{N}}

\def\CP{\mathcal{P}}
\def\CV{\mathcal{V}}
\def\CW{\mathcal{W}}

\def\wfr{\mathfrak{w}}
\def\qfr{\mathfrak{q}}

\def\lgb{\lambda_{\scriptscriptstyle GB}}
\def\ggb{\gamma_{\scriptscriptstyle GB}}
\def\Ngb{N_{\scriptscriptstyle GB}}

\def\pc{\CP_c}

\title{On the connection between hydrodynamics and quantum chaos in holographic theories with stringy corrections}

\author{Sa\v{s}o Grozdanov}
\affiliation{Center for Theoretical Physics, Massachusetts Institute of Technology, \\ Cambridge, MA 02139, USA}
\emailAdd{saso@mit.edu}

\preprint{MIT-CTP/5085}

\abstract{Pole-skipping is a recently discovered signature of many-body quantum chaos in collective energy dynamics. It establishes a precise connection between resummed, all-order hydrodynamics and the underlying microscopic chaos. In this paper, we demonstrate the existence of pole-skipping in holographic conformal field theories with higher-derivative gravity duals. In particular, we first consider Einstein-Hilbert gravity deformed by curvature-squared ($R^2$) corrections and then type IIB supergravity theory with the $\alpha'^3 R^4$ term, where $\alpha'$ is set by the length of the fundamental string. The former case allows us to discuss the effects of leading-order $1/N_c$ corrections (with $N_c$ being the number of colours of the dual gauge group) and phenomenological coupling constant dependence. In Einstein-Gauss-Bonnet theory, pole-skipping turns out to be valid non-perturbatively in the Gauss-Bonnet coupling. The $\alpha'^3 R^4$ deformation enables us to study perturbative inverse 't Hooft coupling corrections ($\alpha'^3 \sim 1 / \lambda^{3/2}$) in $SU(N_c)$, $\mathcal{N} = 4$ supersymmetric Yang-Mills theory with infinite $N_c$. While the maximal Lyapunov exponent characterising quantum chaos remains uncorrected, the butterfly velocity is shown to depend both on $N_c$ and the coupling. Several implications of the relation between hydrodynamics and chaos are discussed, including an intriguing similarity between the dependence of the butterfly velocity and the ratio of shear viscosity to entropy density on stringy corrections.}

\begin{document} 
\maketitle
\flushbottom

\section{Introduction and review of pole-skipping}\label{sec:Intro}

Over the past two decades, our understanding of strongly coupled thermal non-Abelian gauge theories has been greatly advanced by the invention of gauge-string duality (holography) \cite{Maldacena:1997re}. Starting from \cite{Policastro:2001yc,Policastro:2002se}, the duality has helped us paint a robust picture of transport at low energies owing to the fact that strongly coupled states are very well described by hydrodynamics. In other words, real-time thermal correlation functions of holographic theories are largely dominated by hydrodynamic poles in the infra-red part of the spectrum. 

Across families of holographic theories, various detailed properties of hydrodynamic modes have been found to exhibit universal behaviour. In all known cases, holographic statements of universality result from finding boundary (field theory) observables, which can be related in some way to a radially conserved current and computed through only the bulk physics evaluated at the horizon. The most prominent case is the universality of the ratio of shear viscosity to entropy density in first-order hydrodynamics, $\eta/s = 1 / (4\pi)$ \cite{Kovtun:2003wp}. This statement is valid in translationally invariant, two-derivative bulk theories \cite{Kovtun:2004de,Buchel:2003tz,Son:2007vk,Iqbal:2008by,Starinets:2008fb}. In $\CN = 4$ supersymmetric Yang-Mills (SYM) theory at an infinite number of colours, $N_c\to\infty$, the expression is corrected at the leading order in the inverse 't Hooft coupling expansion $\lambda$ \cite{Buchel:2004di,Buchel:2008sh}, i.e. at order $\alpha'^3 \sim 1/\lambda^{3/2}$. In second-order hydrodynamics, there exists a universal (linear) combination of transport coefficients, $H = 2\eta \tau_\Pi - 4 \lambda_1 - \lambda_2$, known as the Haack-Yarom relation \cite{Haack:2008xx}. In analogy with $\eta/s$, $H = 0$ in all holographic two-derivative theories. Moreover, the statement of $H = 0$ remains true even when the leading-order $O(1/\lambda^{3/2})$ coupling corrections are included in $\CN = 4$ SYM theory \cite{Grozdanov:2014kva}, and similarly in curvature-squared bulk theories \cite{Shaverin:2012kv,Grozdanov:2014kva,Grozdanov:2015asa}. However, in Einstein-Gauss-Bonnet theory, where $H$ can be computed non-perturbatively in the higher-derivative coupling, the relation $H=0$ ceases to hold beyond the linear order in the coupling \cite{Grozdanov:2015asa,Grozdanov:2014kva}. As was demonstrated in Refs. \cite{Grozdanov:2015asa,Grozdanov:2014kva} and verified in \cite{Shaverin:2015vda}, perturbatively, the Haack-Yarom relation is violated at second order in $\lgb$, i.e. at $O(\lgb^2)$, where $\lgb$ is the coefficient of the $R^2$ Gauss-Bonnet term (see Ref. \cite{Grozdanov:2016fkt}). In examples of two-derivative non-conformal theories, the validity of $H=0$  was established in \cite{Bigazzi:2010ku,Kleinert:2016nav}. Similar universal relations in third-order hydrodynamics \cite{Grozdanov:2015kqa}, or at higher orders, are so far unknown. In the strongest universal statement to date, by following the observation of \cite{Gursoy:2014boa} that the anomalous chiral magnetic conductivity can also be computed from a horizon formula, Ref. \cite{Grozdanov:2016ala} proved the universality of four anomalous chiral conductivities to all orders in the inverse coupling expansion, i.e. in any classical higher-derivative bulk theory, given certain conditions specified in \cite{Grozdanov:2016ala}. 

Recently, a new type of a universal statement connecting hydrodynamics to the underlying microscopic properties transpired from studies in holography and effective field theory \cite{Grozdanov:2017ajz,Blake:2017ris,Blake:2018leo}. Namely, what was discovered was a new phenomenon termed {\em pole-skipping}, which provides a precise relation between hydrodynamics and many-body quantum chaos. 

To review the phenomenon of pole-skipping and its implications on hydrodynamics and chaos, let us, as in \cite{Grozdanov:2017ajz}, first consider its appearance in $\CN = 4$ SYM theory at infinite 't Hooft coupling $\lambda$, infinite $N_c$ and finite temperature $T$. The physics of the energy-momentum sector in this best understood holographic theory is dual to gravitational dynamics described by the five-dimensional Einstein-Hilbert action with a negative cosmological constant. The background solution dual to a thermal state is the AdS-Schwarzschild black brane. In the longitudinal (sound) channel, the spectrum contains a gapless mode $\omega(k)$---the hydrodynamic sound mode---which in the hydrodynamic gradient expansion of small $|\omega|/T \sim |k|/T \ll 1$ takes the form of an infinite series:
\begin{align}\label{N4Sound}
\omega_\pm (k) = \pm \sum_{n=0}^\infty \CV_{2n+1} k^{2n+1} - i \sum_{n=0}^\infty \Gamma_{2n+2} k^{2n+2}, 
\end{align}
with $\CV_{2n+1} \in \mathbb{R}$ and $\Gamma_{2n+2} \in \mathbb{R}$, for all $n\in \{0\} \cup \mathbb{Z}^+$. $\CV_{2n+1}$ and $\Gamma_{2n+2}$ depend on the all-order hydrodynamic transport coefficients. More precisely, given an integer $m$, $\CV_m$ or $\Gamma_m$ depend on transport coefficients up to order ($m-1$) in the gradient expansion. At present, only $\CV_1 = 1/\sqrt{3}$, $\Gamma_2 = 1/(6\pi T)$,
$\CV_3 = (3-2\ln 2)/(24\sqrt{3}\pi^2 T^2)$ and $\Gamma_4 = (\pi^2 - 24
+ 24\ln 2 - 12 \ln^2 2) / (864 \pi^3 T^3)$ are known analytically---i.e. up to third order in the hydrodynamic expansion, which was constructed and classified in \cite{Grozdanov:2015kqa}.\footnote{See also Refs. \cite{Bu:2014sia,Bu:2014ena}} From the point of view of hydrodynamics, the full gapless quasinormal mode $\omega(k)$ must be thought of as an analytically continued function that arises from a resummation of the entire hydrodynamic series \eqref{N4Sound}, defining $\omega(k)$ for all $k\in\mathbb{C}$.

The statement that relates hydrodynamics to quantum chaos is then that $\omega(k)$---the full gapless quasinormal mode corresponding to the hydrodynamic mode with an all-order, resummed dispersion relation---passes through the following point:
\begin{align}\label{PC}
\CP_c: && \omega(k = i k_0) =i \lambda_L , && \lambda_L = 2\pi T, && k_0 = \lambda_L / v_B. 
\end{align}  
We will refer to this point as the ``point of chaos" because $\lambda_L$ and $v_B$ are the Lyapunov exponent and the butterfly velocity of the underlying quantum chaotic dynamics normally computed by the out-of-time-ordered correlation function (see e.g. Ref. \cite{Maldacena:2015waa}):
\begin{align}\label{OTOC}
C(t,\vecb x) \sim e^{ \lambda_L \left( t -  |\vecb x|/ v_B    \right) }.
\end{align}
In $\CN = 4$ SYM theory at infinite coupling and infinite $N_c$, 
\begin{align}
k_0 = \sqrt{6} \pi T .
\end{align}
The defining property of the point of chaos, which enables one to identify it unambiguously, is that at $\pc$, the longitudinal retarded thermal two-point function of the stress-energy tensor, e.g. the energy density two-point function $G^R_{T^{00} T^{00}}(\omega, k)$, exhibits a very special feature. While it is clear that at $\pc$, $G^R_{T^{00} T^{00}}(\omega = i\lambda_L, k = i k_0)$ has a pole, $G^R_{T^{00} T^{00}}(\omega = i\lambda_L, k = i k_0)$ also has a zero at the same point. Schematically,
\begin{align}\label{EE}
G^R_{T^{00} T^{00}}(\omega, k ) = \frac{b(\omega,k)}{a(\omega ,k )}, && \lim_{(\omega,\,k)\,\to\,\pc} a(\omega, k) = \lim_{(\omega,\,k)\,\to\,\pc} b(\omega,k) = 0.
\end{align}
This is the phenomenon of pole-skipping \cite{Grozdanov:2017ajz,Blake:2017ris,Blake:2018leo}. Since this point defines the value of $\omega$ and $k$ at which the correlator has a pole and a zero, it allows us to find both $\lambda_L$ and $v_B$ without the calculation of the four-point function. One should note that at the point of chaos, both $\omega$ and $k$ are purely imaginary. Hence, recovering this property requires an analytic continuation of the retarded correlator $G^R_{T^{00} T^{00}}(\omega, k ) $ and the dispersion relation $\omega(k)$ to imaginary momentum $k$. 

If, for some reason, the infinite hydrodynamic series truncates at first order in the gradient expansion, then it is possible to directly relate $v_B$ to the thermal diffusion constant. In such cases, one can establish a precise connection between the presently discussed phenomenon and earlier observations of Refs. \cite{Blake:2016sud,Blake:2017qgd,Blake:2016jnn} (see also Ref. \cite{Blake:2018leo}).

In holography, the phenomenon of pole-skipping expressed through Eqs. \eqref{PC} and \eqref{EE} was first observed in the theory discussed above---$\CN = 4$ SYM theory at infinite coupling and infinite $N_c$ \cite{Grozdanov:2017ajz}. This theory at finite coupling and other conformal theories, which posses bulk duals that are deformations of the holographic dual of $\CN = 4$ SYM theory will be of main interest in this work. The fact that relations \eqref{PC} and \eqref{EE} hold generically for holographic theories with two-derivative Einstein-Hilbert gravity coupled to matter was established in \cite{Blake:2018leo}. We will return to a discussion of Ref. \cite{Blake:2018leo} below. From a different point of view, the statements of Eqs. \eqref{PC} and \eqref{EE} also transpired from a (quantum) hydrodynamic effective field theory of chaos proposed in Ref. \cite{Blake:2017ris}. This effective theory describes the dynamics of a single gapless mode, which is responsible for the dynamics of both energy and chaos. A central feature of \cite{Blake:2017ris} is a microscopic reason for pole-skipping---namely, the shift symmetry. In fact, it is this underlying symmetry which acts on the hydrodynamic fields and enforces the hydrodynamic mode $\omega(k)$ to pass through $\pc$, as defined in Eq. \eqref{PC}, and causes $G^R_{T^{00} T^{00}}(\omega, k)$ to have both a pole and a zero at $\pc$, cf. Eq. \eqref{EE}. We note that beyond the effective field theory of \cite{Blake:2017ris}, another non-holographic example of pole-skipping was seen in the study of two-dimensional conformal field theories in the limit of large central charge \cite{Haehl:2018izb}.\footnote{See also Ref. \cite{Cotler:2018zff}.}  

In the context of holography, a gravitational explanation of pole-skipping was provided by Ref. \cite{Blake:2018leo} where it was shown that the phenomenon relates hydrodynamics and chaos in a large class of holographic theories. Namely, it was argued that, generically, in any holographic theory containing two-derivative Einstein gravity coupled to matter (provided sufficient regularity of the bulk matter stress-energy tensor at the horizon) in any spacetime dimension, a quasinormal mode in the sound channel passes through the point of chaos, $\pc$, specified by Eq. \eqref{PC}. In terms of hydrodynamics, this happens regardless of whether the mode is a sound mode or a diffusive mode (in theories with broken translational invariance). Moreover, the energy density correlator exhibits pole-skipping \eqref{EE} with both a zero and a pole of $G^R_{T^{00} T^{00}}(\omega, k)$ passing through $\pc$. In the spirit of all universal holographic statements, the arguments of \cite{Blake:2018leo} also relate the phenomenon of pole-skipping and the value of $\pc$ to bulk dynamics evaluated purely at the black hole horizon. The reasons for this stem from an identification of a new type of behaviour of (horizon) Einstein's equations. By writing the gravitational equations of motion in terms of ingoing Eddington-Finkelstein coordinates, with $v$ the ingoing null coordinate, it turns out that the $vv$ component of Einstein's equations identically vanishes at the horizon when $\omega$ and $k$ take the imaginary values of the chaos point $\pc$. That is, $E_{vv} = 0$, identically, which allows one to infer the existence of pole-skipping. Furthermore, this property even allows one to compute the slope of the dispersion relations $\omega(k)$ at $\pc$. For details, see Ref. \cite{Blake:2018leo}.  

In this work, we will extend the holographic discussions of \cite{Grozdanov:2017ajz} and \cite{Blake:2018leo} to examine whether the above-presented relation between hydrodynamics and quantum chaos can exist in the presence of stringy bulk corrections---i.e. in field theories at finite 't Hooft coupling $\lambda$ and finite $N_c$. In Sections \ref{sec:GB} and \ref{sec:N4}, we will show that to leading order in $1/N_c$ and inverse coupling, this is indeed the case. Moreover, we will show that pole-skipping remains valid even for non-perturbative values of the curvature-squared coupling in Einstein-Gauss-Bonnet theory. Finally, in Section \ref{sec:Discussion}, we will discuss various physical implications of these results with the aim of establishing in greater generality the fascinating fact that quantitative characteristics of microscopic quantum chaos can be inferred from macroscopic, collective hydrodynamics.

\section{Curvature-squared theories}\label{sec:GB}

Let us begin our study of pole-skipping in strongly coupled, non-Abelian large-$N_c$ conformal field theories (CFTs) at finite temperature with holographic higher-derivative gravity duals. Throughout this work, we will think of $SU(N_c)$, $\CN = 4$ supersymmetric Yang-Mills (SYM) theory as our central example. The energy-momentum dynamics of a thermal state in $\CN =4$ SYM theory at $N_c \to \infty$ and $\lambda\to\infty$ is encoded in the dual Einstein-Hilbert action, 
\begin{align}
S = \frac{1}{2 \kappa_5^2 } \int d^5 x \sqrt{-g} \left( R - 2 \Lambda \right),
\label{eq:N4}
\end{align}
with a negative cosmological constant $\Lambda = - 6 / L^2$. We set the AdS radius to $L=1$. The five-dimensional Newton's constant is related to $N_c$ via $\kappa_5 = 2\pi /N_c$. We first consider the family of curvature-squared deformations of the action \eqref{eq:N4}:
\begin{align}
S_{R^2} = \frac{1}{2 \kappa_5^2 } \int d^5 x \sqrt{-g} \left( R +12 + \alpha_1 R^2 + \alpha_2 R_{\mu\nu} R^{\mu\nu} + \alpha_3 R_{\mu\nu\rho\sigma} R^{\mu\nu\rho\sigma} \right).
\label{eq:R2Th}
\end{align}
Since \eqref{eq:R2Th} generates higher-than-second-order equations of motion, the three coupling constants $\alpha_i$ should be thought of as perturbatively small, $|\alpha_i| \ll 1$. It is well known that the couplings $\alpha_1$ and $\alpha_2$ can be removed via a field redefinition of the metric tensor at the expense of introducing $O(\alpha^2)$ corrections (see \cite{Brigante:2007nu,Buchel:2008vz,Grozdanov:2014kva}). For this reason, the results that follow from the action \eqref{eq:R2Th} with $\alpha_3 = 0$ can be directly inferred from those computed from the pure Einstein-Hilbert action \eqref{eq:N4} dual to $\CN = 4$ SYM theory at infinite coupling $\lambda$ and infinite $N_c$ after performing simple parameter redefinitions and constant Weyl transformations (see \cite{Grozdanov:2014kva} and also \cite{Grozdanov:2016vgg}). 

It is convenient to use this field redefinition to set $\alpha_1$ and $\alpha_2$ in a way that turns the $R^2$ part of the action into the Gauss-Bonnet term: 
\begin{align}\label{GBaction}
S_{GB} = \frac{1}{2\kappa_5^2} \int d^5 x \sqrt{-g} \left[ R + 12 + \frac{\lgb}{2} 
\left( R^2 - 4 R_{\mu\nu} R^{\mu\nu} + R_{\mu\nu\rho\sigma} R^{\mu\nu\rho\sigma} \right) \right].
\end{align}
A particularly useful feature of this action is that it produces equations of motion which contain derivatives only up to second order. Thus, $\lgb$ can be formally treated as a non-perturbative coupling.  For a detailed discussion regarding various aspects of this theory, see Ref. \cite{Grozdanov:2016fkt}. We will neglect any potential subtleties that may arise from an inclusion of a stringy tower of massive higher-spin states, which is necessary to cure the ultra-violet problems of curvature-squared theories \cite{Camanho:2014apa}.

From the point of view of string theory, there are two ways to interpret the coupling $\lgb$. It can be considered as tuning the dual coupling constant $\lambda$---a view which is reinforced by the fact that, for example, in bosonic and heterotic string theory, $\lgb$ is proportional to $\alpha'$. A multitude of predictions that follow from Einstein-Gauss-Bonnet theory for negative values of $\lgb$ is consistent with qualitative expectations of the way thermal physics should transition from strong to weak coupling \cite{Grozdanov:2015asa,Grozdanov:2016vgg,Grozdanov:2016fkt,Grozdanov:2016zjj,Andrade:2016rln,Atashi:2016fai,DiNunno:2017obv,Grozdanov:2018fic}. Another way to interpret $|\lgb| \ll 1$ is through the fact that the four-dimensional dual of \eqref{GBaction} has conformal central charges $c$ and $a$ which are no longer equal. There exists a parameter regime, e.g. $1 \ll N_c^{2/3} \ll \lambda$, in which $\alpha_3 \sim \lgb $ can be associated with $1/N_c \ll 1$ corrections so that $(c-a)/c \sim 1/N_c$. A valuable example of such a theory, which we will make use of in this note, is $Sp(N_c)$, $\CN=2$ superconformal gauge theory studied by Kats and Petrov in \cite{Kats:2007mq}. In string theory, this theory arises from a construction of $N_c$ D3-branes embedded inside 8 D7-branes coincident on an orientifold 7-plane. Its supergravity dual descends from type IIB string theory on AdS${}_5 \times X^5$, where $X^5 \simeq S^5 / \mathbb{Z}_2$ (see Ref. \cite{Aharony:1998xz}). By computing the central charges from field theory, where $c = \frac{1}{24}(12 N_c^2+18 N_c-2)$ and $a = \frac{1}{24} (12 N_c^2 + 12 N_c - 1 )$, and holography, it follow that $\alpha_3 = (c-a)/8c$. Written in terms of the Gauss-Bonnet coupling,  
\begin{align}\label{SpLgb}
\lgb =2 \alpha_3 =  \frac{1}{8 N_c}.
\end{align}
Such corrections do not incorporate any stringy higher-order worldsheet genus corrections, i.e. corrections in the string coupling $g_s$.\footnote{For discussions of (quantum) $1/N_c^2$ effects in holography, see \cite{Denef:2009yy,Denef:2009kn,CaronHuot:2009iq,Ardehali:2013gra,Ardehali:2013xya,Arnold:2016dbb}.} Generically, we expect $\lgb$ (or $\alpha_i$) to be a function of both $N_c$ (or $g_s$) and $\lambda$ (or $\alpha'$). For a more detailed discussion of these issues in relation to $R^2$ theories, see \cite{Buchel:2008vz} and also \cite{Grozdanov:2016fkt}.

After this interlude, we return to studying pole-skipping in Einstein-Gauss-Bonnet theory. We will do this non-perturbatively in $\lgb$, with the understanding that to linear order in $\lgb$ (or $\alpha_i$), the theories in Eqs. \eqref{eq:R2Th} and \eqref{GBaction} are equivalent. The second-order equations of motion that follow from the action \eqref{GBaction} are 
\begin{align}\label{GB-EOM}
E_{\mu\nu} \equiv& \; R_{\mu\nu} - \frac{1}{2} g_{\mu\nu} R + g_{\mu\nu} \Lambda - \frac{\lgb}{4} g_{\mu\nu} \left( R^2 - 4 R_{\mu\nu} R^{\mu\nu} + R_{\mu\nu\rho\sigma} R^{\mu\nu\rho\sigma}  \right) \nn  
&+ \lgb  \left( R R_{\mu\nu} - 2 R_{\mu\alpha} R_{\nu}^{~\alpha} - 2 R_{\mu\alpha\nu\beta} R^{\alpha\beta} + R_{\mu\alpha\beta\gamma} R_\nu^{\alpha\beta\gamma}  \right)  =  0.
\end{align}
In ingoing Eddington-Finkelstein coordinates, the black brane metric is
\begin{align}
ds^2 &= - f(r) \Ngb^2 dv^2 + 2 \Ngb dv dr + r^2 d\vecb{x}^2, \\
f(r) &=  \frac{r^2}{2\lgb} \left[1 - \sqrt{1-4\lgb \left(1 - \frac{r^4_0}{r^4} \right) } \right],
\end{align}
where $r_0$ is the position of the horizon and the Hawking temperature is $T = \Ngb r_0 / \pi$. The constant $\Ngb$ is chosen so as to ensure that the boundary speed of light equals to one:
\begin{align}
\Ngb^2 =  \frac{1}{2} \left( 1 + \sqrt{1-4\lgb} \right).
\end{align}
For future convenience, we also define $\ggb \equiv \sqrt{1 - 4\lgb}$. In the sound channel, the hydrodynamic dispersion relation is known to second order in the gradient expansion, i.e. to order $k^3$ (see Refs. \cite{Grozdanov:2016vgg,Grozdanov:2016fkt}),  
\begin{align}
\wfr =& \pm \frac{1}{\sqrt{3}} \qfr - \frac{1}{3} i \ggb^2\qfr^2  \nn
& \mp  \frac{1}{12 \sqrt{3}} \ggb  \left(2 + \ggb ^3-6 \ggb ^2-3 \ggb +2 \ggb  \ln \left[ \frac{2 (1+\ggb )}{\ggb }\right] \right)  \qfr^3 + \cdots \, , \label{sound-qnm}
\end{align}
where $\wfr \equiv \omega / (2\pi T)$ and $\qfr \equiv k/(2 \pi T)$.

\begin{figure}[t]
\begin{center}
\includegraphics[width=0.66\textwidth]{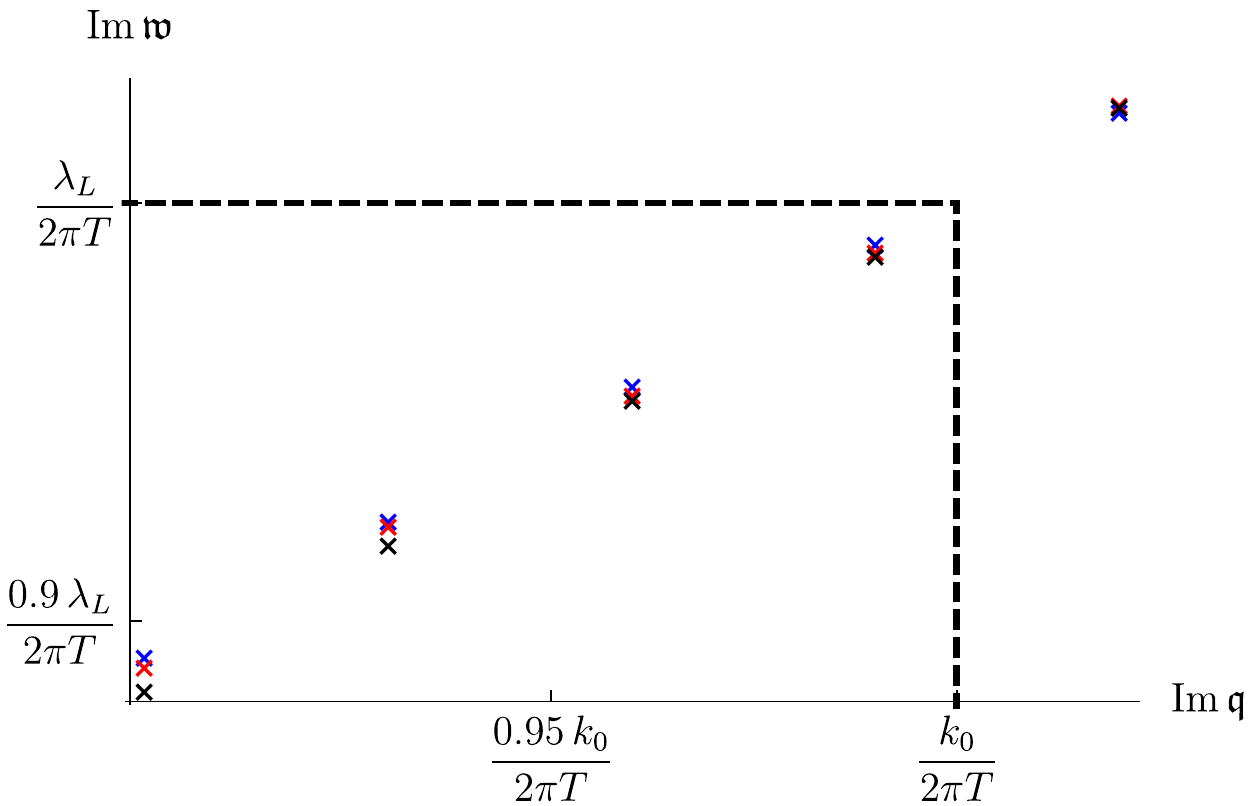}
\caption{A plot of the hydrodynamic sound dispersion relation $\wfr(\qfr)$ in non-perturbative Einstein-Gauss-Bonnet theory for a choice of three different values of $\lgb = \{1/8, -1,-4\}$ depicted in black, blue and red, respectively. We see that for all $\lgb$, the dispersion relation curves pass through the point of chaos \eqref{ChaosPointGB}, as enforced by the phenomenon of pole-skipping.}
\label{fig:GB}
\end{center}
\end{figure}

In Einstein-Gauss-Bonnet theory, the arguments of \cite{Blake:2018leo} can be repeated verbatim. In ingoing EF coordinates, the horizon value of the $E_{vv}$ component of equations \eqref{GB-EOM} is
\begin{align}
E_{vv} = \left(- \frac{3 i \pi T}{\Ngb^2} \omega  + k^2 \right)\Ngb \delta g_{vv}^{(0)} - i \Ngb \left(2\pi T + i \omega\right) \left[ \omega \delta g^{(0)}_{x^i x^i} + 2 k \delta g^{(0)}_{vx}   \right] = 0.
\end{align}
Hence, $E_{vv} = 0$, identically, when $\omega$ and $k$ take the imaginary chaos point values $\pc$:
\begin{align}\label{ChaosPointGB}
\omega = i\lambda_L =  2\pi T i , && k = i k_0, && k_0 = \frac{\sqrt{6} \pi T }{\Ngb},
\end{align}
with the butterfly velocity
\begin{align}
v_B = \frac{\lambda_L}{k_0} = \sqrt{\frac{2}{3}} \left(\frac{1 + \ggb}{2}   \right)^{1/2}  . 
\end{align}
Both $\lambda_L$ and $v_B$ precisely match the results found from the shock wave calculation of Ref. \cite{Roberts:2014isa}, non-perturbatively in the $\lgb$ coupling.\footnote{See also Refs. \cite{Alishahiha:2016cjk,Mezei:2016wfz}. For a calculation of $v_B$ in three-dimensional gravity with an $R^2$ correction, see \cite{Qaemmaqami:2017jxz}.} By following \cite{Blake:2018leo}, we can infer that \eqref{ChaosPointGB} gives the imaginary values of frequency and momentum at which the energy density two point functions has a zero and a pole---i.e., it is the point of pole-skipping. We numerically verify that the pole that passes through the point of chaos is again the hydrodynamic sound mode (see Figure \ref{fig:GB}). Finally, due to the field redefinition discussed below Eq. \eqref{eq:R2Th}, the fact that pole-skipping exists for non-perturbative values of $\lgb$ implies that it also occurs perturbatively in $\alpha_i$ in the most general $R^2$ theory \eqref{eq:R2Th}. 

To connect this calculation with the question of the existence of pole-skipping in field theories with parameter regimes of $N_c$ and $\lambda$ away from infinity, we return to the example of $Sp(N_c)$, $\CN =2$ gauge theory. Using Eq. \eqref{SpLgb}, the leading $1/N_c$-dependent correction to the butterfly velocity becomes
\begin{align}
v_B =  \sqrt{\frac{2}{3}} \left(1 - \frac{\lgb}{2 } + \cdots \right) = \sqrt{\frac{2}{3}} \left(1 - \frac{1}{16 N_c } + \cdots \right).
\end{align}
Thus, $v_B$ decreases as a consequence of the leading-order $1/N_c$ correction. In many other examples of theories collected in \cite{Buchel:2008vz}, the butterfly velocity also becomes corrected by a term with a negative sign so long as the perturbative effects of the $R^2$ terms are interpreted through the difference between the two central charges of the CFT, $(c-a)/c$. That is, $v_B = \sqrt{2/3} \left(1 + \delta\right)$, with $\delta$ negative, $-1 \ll \delta <0$. This is in contrast with the phenomenological interpretation of $\lgb$ as the coupling constant correction, i.e. with $\lgb < 0$ being the natural sign \cite{Grozdanov:2015asa,Grozdanov:2016vgg,Grozdanov:2016fkt,Grozdanov:2016zjj,Andrade:2016rln,Atashi:2016fai,DiNunno:2017obv,Grozdanov:2018fic}, under which $v_B$ increases ($\delta$ is positive). 

\section{$\CN = 4$ supersymmetric Yang-Mills theory at finite 't Hooft coupling}\label{sec:N4}

We now turn our attention to a top-down (type IIB supergravity theory) study of coupling constant corrections to pole-skipping in $\CN=4$ SYM theory in the limit of an infinite number of colours, $N_c\to \infty$. To study the energy-momentum sector of this strongly coupled CFT in a perturbative inverse 't Hooft coupling expansion around $\lambda \to\infty$, we need to consider the leading-order, higher-derivative ($\alpha'^3 R^4$) correction to the Einstein-Hilbert action:
\begin{align}
S = \frac{1}{2\kappa_5^2} \int d^5 x \sqrt{-g} \left(R  + \frac{12}{L^2} + \gamma \CW \right),
\label{eq:hd-action}
\end{align}
where in terms of the Weyl tensor $C_{\alpha\beta\gamma\delta}$, 
\begin{align}
\label{eq:Wterm}
\CW = C^{\alpha\beta\gamma\delta}C_{\mu\beta\gamma\nu} C_{\alpha}^{~\rho\sigma\mu} C^{\nu}_{~\rho\sigma\delta} + \frac{1}{2} C^{\alpha\delta\beta\gamma} C_{\mu\nu\beta\gamma} C_{\alpha}^{~\rho\sigma\mu} C^\nu_{~\rho\sigma\delta}.
\end{align}
The action \eqref{eq:hd-action} arises from a dimensional reduction of the ten-dimensional type IIB supergravity action on $S^5$. The perturbative higher-derivative coupling $\gamma = \alpha'^3 \zeta(3)/8 $ is related to the dual 't Hooft coupling via $\gamma  = \lambda^{-3/2}\zeta (3) L^6/8$. Again, $L$ is set to $L=1$ so that $\alpha'  = 1/\sqrt{\lambda}$. The relation between the five-dimensional gravitational constant and $N_c$ is the same as in $\CN = 4$ SYM theory at infinite coupling, i.e. $\kappa_5 = 2\pi /N_c$. For other details regarding the theory, see \cite{Grozdanov:2014kva,Grozdanov:2016vgg}.

In ingoing EF coordinates ($v,r,x^i$), with
\begin{align}\label{EFN4}
v = t +r_*(r), && \frac{ dr_*}{dr} = \frac{1}{r^2 f(r)} \frac{Z_{vr}(r)}{Z_{vv}(r)},
\end{align}
the five-dimensional $\gamma$-corrected black brane metric of \cite{Gubser:1998nz,Pawelczyk:1998pb}, which is a solution to the equations of motion derived from \eqref{eq:hd-action}, takes the form
\begin{align}\label{N4BBEF}
ds^2 = - r^2 f(r) Z_{vv}(r) dv^2 + 2 Z_{vr} (r) dv dr + r^2 d\vecb{x}^2.
\end{align}
The functions appearing in the line element are
\begin{align}
f(r)&= 1 - \left( \frac{r_0}{r} \right)^4,\label{fNlambda} \\ 
Z_{vv}(r) &= 1 - \gamma \left[ 75\left( \frac{r_0}{r} \right)^4 + 75 \left( \frac{r_0}{r} \right)^8 - 45 \left( \frac{r_0}{r} \right)^{12}    \right], \label{Zvv} \\
Z_{vr}(r) &= 1 - 120 \gamma  \left( \frac{r_0}{r} \right)^{12} ,\label{Zvr}
\end{align}
and the Hawking temperature is 
\begin{align}\label{T-N4}
T = \frac{r_0 }{ \pi} (1 + 15 \gamma).
\end{align} 
In the presence of $\gamma$-dependent corrections, the hydrodynamic sound mode dispersion relation is known to second order in the gradient expansion (see Refs. \cite{Grozdanov:2014kva,Grozdanov:2016vgg}):
\begin{align}
\wfr = \pm \frac{1}{\sqrt{3}} \qfr - \frac{1}{3} i \left(1 + 120\gamma\right) \qfr^2 \mp \left[ \frac{-3 + 2\ln 2}{6\sqrt{3} }+ \frac{5 \left(-41 + 16 \ln 2 \right)}{2\sqrt{3}} \gamma   \right]\qfr^3  + \cdots.
\end{align}
The complete list of $\gamma$-corrected hydrodynamic first- and second-order transport coefficients can be found  in \cite{Grozdanov:2014kva}. The (third-order) hydrodynamic coefficient of $\qfr^4$ is only known in the limit of $\lambda \to \infty$ (or $\gamma \to 0$) \cite{Grozdanov:2015kqa}.

Repeating the same calculation as in Section \ref{sec:GB} (or \cite{Blake:2018leo}) to first order in a perturbative expansion around infinite 't Hooft coupling, i.e. by expanding the metric components as
\begin{align}
\delta g^{(n)}_{\mu\nu} = \delta g^{(n,0)}_{\mu\nu}+\gamma \delta g^{(n,1)}_{\mu\nu},
\end{align}
the $vv$ component of Einstein's equations takes the schematic form
\begin{align}\label{EvvN4}
E_{vv} = \left(- 3 i \pi T \omega  + k^2 \right) \delta g_{vv}^{(0,0)} - i \left(2\pi T + i \omega\right) \left[ \omega \delta g^{(0,0)}_{x^i x^i} + 2 k \delta g^{(0,0)}_{vx}   \right] + \gamma E_{vv}^{(1)} = 0.
\end{align}
$E_{vv}^{(1)}$ is given by an unilluminating lengthy expression, which we do not state here. In order to find $\omega$ and $k$ such that $E_{vv} = 0$, we write
\begin{align}
\omega = 2 \pi T i \left( 1 + \omega_1 \gamma \right), && k = \sqrt{6} \pi T i \left(1 + k_1 \gamma \right).
\end{align}
Eq. \eqref{EvvN4} then reduces to
\begin{align}
E_{vv} = \gamma E_{vv}^{(1)}  = 2 \pi^2 T^2 \gamma \left[ 3 \left(\omega_1 - 2 k_1 - 23 \right) \delta g_{vv}^{(0,0)} - 2 \omega_1 \left( \delta g^{(0,0)}_{x^i x^i} + \sqrt{6}\, \delta g^{(0,0)}_{vx}  \right)   \right] = 0,
\end{align}
which implies that $E_{vv} = 0$, identically, if and only if $\omega_1 = 0$ and $k_1 = - 23/2$. That is, 
\begin{align}
\omega = 2 \pi T i , && k = i k_0 = \sqrt{6} \pi T i \left(1 - \frac{23}{2} \gamma \right) = \sqrt{6} \pi T i \left(1 - \frac{23 \zeta(3) }{16} \frac{1}{\lambda^{3/2}}   \right).
\end{align}
In other words, pole-skipping occurs at the point $\pc$ where $\omega$ is fixed by the maximal holographic Lyapunov exponent $\lambda_L$ while the ratio of $\omega/k$ predicts a coupling-dependent butterfly velocity $v_B$: 
\begin{align}\label{PCGamma}
\lambda_L = 2\pi T, && v_B = \frac{\lambda_L}{k_0} = \sqrt{ \frac{2}{3} }\left(1 + \frac{23 \zeta(3) }{16} \frac{1}{\lambda^{3/2}}  \right) .
\end{align}
We note that $v_B$ increases as $\lambda$ is tuned from infinite towards intermediate coupling. This qualitative feature is yet again \cite{Grozdanov:2015asa,Grozdanov:2016vgg,Grozdanov:2016fkt,Grozdanov:2016zjj,Andrade:2016rln,Atashi:2016fai,DiNunno:2017obv,Grozdanov:2018fic} consistent with the negative sign of $\lgb$ in Einstein-Gauss-Bonnet theory (cf. Section \ref{sec:GB}) being interpreted as the coupling correction rather than a correction in $1/N_c$.  

Expression \eqref{PCGamma} is a prediction for the behaviour of microscopic many-body chaos and its effect on the hydrodynamic sound mode away from infinite coupling. It is particularly valuable because studying properties of scrambling in $\CN = 4$ SYM theory at finite $\lambda$ is a much more involved problem than at infinite $\lambda$ or in the case of $R^2$ theories (cf. Section \ref{sec:GB}). Stringy effects on scrambling in holography were first considered by Shenker and Stanford in Ref. \cite{Shenker:2014cwa}. Their string-corrected scattering phase factor, which controls the growth of the OTOC, was given by  
\begin{align}\label{StringShock}
\delta (t,\vecb x) = 4\pi G_N s(t) \int \frac{d^3k}{(2\pi)^3} \frac{e^{i \vecb k\cdot \vecb x}}{k^2 + \mu^2 } \left( \frac{e^{-i\pi/2} s(t) \alpha' }{4}  \right)^{-\alpha' (k^2+\mu^2)/2 r_0^2 } ,
\end{align}
where $k^2$ is the square of the spatial momentum vector, i.e. $k^2 = \vecb k \cdot \vecb k$. In the expression \eqref{StringShock}, $s(t)\sim e^{2\pi T t}$, we used the AdS radius $L=1$ and set the string length to $\ell_s^2 = \alpha' = 1/\sqrt{\lambda} $. In our notation, $\mu = \sqrt{6} r_0$ and $r_0 = \pi T_0$, where $T_0$ is the temperature of the system at infinite coupling ($\alpha' = 1/\sqrt{\lambda} \to 0$). Note that $\mu$ equals to $k_0$, also evaluated at $\alpha' = 1/\sqrt{\lambda} \to 0$. Ref. \cite{Shenker:2014cwa} observed that the integral in \eqref{StringShock} has two distinct regimes in which the OTOC exhibits the following two distinct types of behaviour:
\begin{align}
 t  \ll \sqrt{\lambda} |\vecb x |: && C(t,\vecb{x}) &\sim \frac{e^{2\pi T\left(t - |\vecb x| / v_B \right)} }{|\vecb x | }  , \\
 \sqrt{\lambda} |\vecb x | \ll t: && C(t,\vecb{x}) &\sim  \frac{ e^{2\pi T \left(1 - \frac{3}{ \sqrt{\lambda}} \right) t - \frac{ |\vecb x|^2 }{ 2\rho^2} }}{\rho^3} e^{\frac{i\pi}{2} \frac{3}{\sqrt{\lambda}}},\label{shortdistSSRes}
\end{align}
where $\rho^2 \approx 2 \sqrt{\lambda} t / (\pi T)$. At long distances ($ t  \ll \sqrt{\lambda} |\vecb x |$), the effects of string scattering that are included in recovering \eqref{StringShock} result in exponential growth which is unaffected by $\alpha'$. For this reason, the long-distance behaviour of scrambling is uncorrected as compared to the maximally chaotic, infinitely strongly coupled result:
\begin{align} 
t  \ll \sqrt{\lambda} |\vecb x |: && \lambda_L = 2 \pi T, && v_B = \sqrt{\frac{2}{3}}.
\end{align}
At very short distance scales ($ t  \gg \sqrt{\lambda} |\vecb x |$), the effect of stringy corrections enters at order $\alpha' = 1/\sqrt{\lambda}$ and spreads the shock. The exponential spreading of scrambling has a quadratic dependence on $|\vecb x|$ and the Lyapunov exponent is corrected by a coupling-dependent shift:
\begin{align}\label{SSLambda}
\sqrt{\lambda} |\vecb x | \ll t: && \lambda_L = 2 \pi T \left(1 - \frac{3}{\sqrt{\lambda}} \right).
\end{align}
The concept of the butterfly velocity is no longer meaningful because $v_B$ can only be defined in the region of $t\sim |\vecb x| / \lambda^{1/4}$, which is far from the regime where Eq. \eqref{shortdistSSRes} is valid. It is also worth noting that at short distances, $C(t,\vecb x)$ becomes complex with a (small) $\alpha'$-dependent phase. 

The classical supergravity action \eqref{eq:hd-action} considered in this section gave rise to a predicted $\alpha'^3 = 1 / \lambda^{3/2}$ correction to $k_0$ of pole-skipping, and thereby the butterfly velocity. Since we expect our perturbative $\alpha'^3$-dependent result to be applicable for characterising chaos at long-distance scales, we can determine that Eq. \eqref{StringShock} does not capture this effect. The natural question is therefore whether the classical $\alpha'^3$-corrected shock wave geometry that follows from \eqref{eq:hd-action} and directly probes scrambling reproduces the prediction of Eq. \eqref{PCGamma}. The remaining task is to find such a solution and use the result to determine the dual OTOC via the holographic dictionary of Ref. \cite{Roberts:2014isa}. Then, we will be able to infer the butterfly velocity explicitly from properties of scrambling and test the (hydrodynamic) prediction \eqref{PCGamma} of pole-skipping.

We first write the black brane geometry \eqref{N4BBEF} in terms of Schwarzschild coordinates, 
\begin{align}
ds^2 = - r^2 f(r) Z_{vv}(r) dt^2 + \frac{Z_{vr}(r)^2}{r^2 f(r) Z_{vv}(r)} dr^2 + r^2d \vecb{x}^2 ,
\end{align}
where $f(r)$, $Z_{vv}(r)$ and $Z_{vr}(r)$ can be found in Eqs. \eqref{fNlambda}, \eqref{Zvv} and \eqref{Zvr}. Now, define Kruskal-Szekeres (KS) coordinates:
\begin{align}
U = - e^{-2\pi T u}, && V = e^{2\pi T v},
\end{align}
where $u = t - r_*(r)$ and, as before, $v = t + r_*(r)$ (cf. Eq. \eqref{EFN4}). In terms of these coordinates that enable a maximal extension of the Schwarzschild spacetime,
\begin{align}
ds^2 = A(UV) dUdV + B(UV) d \vecb{x}^2 ,
\end{align}
where
\begin{align}
A(UV) & = \frac{r^2 f(r) Z_{vv}(r)}{\left(2\pi T\right)^2 UV }, \\
B(UV) & = r^2 .  
\end{align}
To construct the Dray-'t Hooft shock wave solution that propagates on top of this background, we perturb the metric along a single spatial component, $\delta ds^2 = - A(UV) h(z) \delta (U) dU^2$, and solve the equations of motion for $h(z)$. In a small $\gamma$ expansion, we write $A(UV)$, $B(UV)$ and $h(z)$ in the following form:
\begin{align}
A(UV) &= A_0(UV) + \gamma A_1(UV) ,\\
B(UV) &= B_0(UV) + \gamma B_1 (UV) ,\\
h(z) &= h_0(z) + \gamma h_1 (z) .
\end{align}
At zeroth order in $\gamma$, the background functions satisfy
\begin{align}
A_0' &= \frac{A_0 \left[ 2 A_0 B_0^2 - B_0'\left(B_0 + UV B_0'\right) \right]}{UV B_0 B_0'} , \\
B_0'' &= \frac{2 B_0 A_0' B_0' + A_0 (B_0')^2}{2A_0 B_0},
\end{align}
where ${}'$ denotes a derivative with respect to $UV$. It is easy to check that the black brane with $\gamma=0$ (AdS${}_5$-Schwarzschild solution) satisfies these relations. We take higher derivatives of these expressions to simplify the $O(\gamma)$ equations. The four-derivative equations of motion at $O(\gamma)$ are rather involved, so we do not state the analogous background relations here. To find the shock wave propagating at $U=0$, it is sufficient to know the values of $A(UV)$ and $B(UV)$ at $UV =0$. Explicitly, 
\begin{align}
A(0) &= A_0(0) + \gamma A_1(0), &&A_0(0) = -\frac{1}{r_0}, && A_1(0) = \frac{135}{r_0} , \\
A'(0) &= A'_0(0) + \gamma A'_1(0), &&A'_0(0) = -\frac{1}{r_0^2}, && A'_1(0) =- \frac{585}{r_0^2} , \\
B(0) &= B_0(0) + \gamma B_1(0), &&B_0(0) = r_0^2, && B_1(0) = 0 , \\
B'(0) &= B'_0(0) + \gamma B'_1(0), &&B'_0(0) = - 2 r_0, && B'_1(0) = 0 , \\
B''(0) &= B''_0(0) + \gamma B''_1(0), &&B''_0(0) = 0, && B''_1(0) = - 4320  ,
\end{align}
along with $A_0''(0) = - \frac{9}{2 r_0^3}$. With these relations in hand, the remaining two equations of motion that need to be solved determine $h_0(z)$ and $h_1(z)$, and thereby the butterfly velocity. At zeroth order in $\gamma$, 
\begin{align}\label{Eq0vB}
\left[ \partial^2_z - 6 r_0^2 \right] h_0 (z) \sim e^{2\pi T t_w} E \delta(z), 
\end{align}
where $E$ characterises the energy of the perturbation that sources the shock wave. At first order, we find a non-homogeneous differential equation
\begin{align}\label{Eq1vB}
\left[ \partial_z^2 - 6 r_0^2 \right] h_1 (z)  = 42 r_0^2  h_0 (z).  
\end{align}
Solving Eqs. \eqref{Eq0vB} and \eqref{Eq1vB}, and choosing $h_0(z)$ and $h_1(z)$ to be exponentially decaying, we find
\begin{align}
h_0(z) &=   e^{-\sqrt{6} r_0  z }, \\
h_1(z) &= - 7  \left( \frac{1}{4} + \frac{1}{2} \sqrt{6} r_0  z \right) e^{-\sqrt{6} r_0 z } + H_1 e^{-\sqrt{6} r_0 z } ,
\end{align}
where $H_1$ in an integration constant. Setting $H_1 = 7 / 4$, and exponentiating the sum of $h_0(z) + \gamma h_1(z)$, we arrive at the solution for $h(z)$ valid up to $O(\gamma)$,
\begin{align}
h(z) = \exp\left\{-\sqrt{6} r_0 \left(1 +  \frac{7}{2}  \gamma \right) z \right\},
\end{align}
or written in terms of the temperature, i.e. through $ r_0 = \pi T (1 - 15 \gamma)$,
\begin{align}
h(z) = \exp\left\{-\sqrt{6} \pi T \left(1 - \frac{23}{2} \gamma \right) z \right\} .
\end{align}
Hence, the OTOC in $\CN = 4$ SYM theory at $N_c \to\infty$, corrected at the order $\alpha'^3 \sim 1 / \lambda^{3/2}$, has the following exponential behaviour:
\begin{align}
C(t,z) \sim e^{\lambda_L \left(t - z / v_B \right)} = e^{2\pi T t} h(z) = \exp\left\{ 2\pi T \left[ t - \sqrt{\frac{3}{2}} \left(1 - \frac{23}{2} \gamma \right) z \right]   \right\} ,
\end{align}
or, in other words, we find
\begin{align}
\lambda_L = 2\pi T, && v_B = \sqrt{ \frac{2}{3} }\left(1 + \frac{23}{2} \gamma \right) = \sqrt{ \frac{2}{3} }\left(1 + \frac{23 \zeta(3) }{16} \frac{1}{\lambda^{3/2}}  \right) ,
\end{align}
which precisely agrees with the pole-skipping prediction for the (maximal) Lyapunov exponent and the coupling-dependent butterfly velocity stated in Eq. \eqref{PCGamma}. 

\section{Discussion}\label{sec:Discussion}

The picture of the connection between hydrodynamics and chaos that is beginning to emerge can be summarised in the following way. Let us first assume that the number of colours is the gauge theory is infinite, $N_c \to \infty$. This enables us to use the OTOC as a good measure of late-time chaos \cite{Kukuljan:2017xag}. At infinite 't Hooft coupling, $\lambda \to \infty$, chaos spreads ballistically with the maximal holographic Lyapunov exponent $\lambda_L = 2 \pi T$ and the butterfly velocity $v_B$. In $\CN = 4$ SYM theory, $v_B = \sqrt{2/3}$. Both $\lambda_L$ and $v_B$ can be inferred from the behaviour of the gapless hydrodynamic mode in the longitudinal (sound) channel at the point of chaos $\pc$ through the phenomenon of pole-skipping \cite{Grozdanov:2017ajz}. As discussed in the Introduction, pole-skipping is a generic phenomenon in duals of bulk two-derivative gravity theories coupled to matter \cite{Blake:2018leo}.  

At infinite $N_c$, but as $\lambda$ is (perturbatively) tuned away from infinity, we begin to observe the effects of $\lambda$-dependent scale separation.\footnote{For discussions of scale separation in low- and high-energy regions of the thermal spectrum of stress-energy tensor correlators, and in hydrodynamics and thermalisation, see \cite{Grozdanov:2016vgg,Solana:2018pbk}.} At very short distances, $\sqrt{\lambda} |\vecb x | \ll t$, the dual bulk description is dominated by high-energy string scattering processes. In consequence, the leading Lyapunov exponent that characterises the exponential growth of chaos is corrected at order $\alpha'=1/\sqrt{\lambda}$ \cite{Shenker:2014cwa}:
\begin{align}
\lambda_L = 2 \pi T \left( 1 - \frac{\delta \lambda^{(1)}_L}{\sqrt{\lambda}} + O\left(1/\lambda\right) \right),
\end{align}
with $\delta \lambda^{(1)}_L > 0$, as per \cite{Maldacena:2015waa}. Because the exponential growth of the OTOC scales as $e^{\lambda_L t - |\vecb x|^2/2\rho^2}$, there is no natural butterfly velocity. Since chaos is characterised by an expression which is an integral over spatial momenta $\vecb k$ (cf. Eq. \eqref{StringShock}), in general, we cannot speak of a single Lyapunov exponent. Rather, one can associate a Lyapunov exponent to each momentum mode by writing the OTOC in the following form: 
\begin{align}
C(t,\vecb x ) \sim \int \frac{d^3k}{(2\pi)^3} e^{i \vecb k\cdot \vecb x}  e^{ \lambda_L(k) t}\Gamma(k), && \lambda_L(k) = 2\pi T \left( 1 - \frac{k^2 + 6 \pi^2 T^2}{2\pi^2 T^2} \frac{1}{\sqrt{\lambda}} \right),
\end{align}
where $\Gamma(k)$ is some non-exponential function. The dominant Lyapunov exponent $\lambda_L$ in \eqref{SSLambda} corresponds to the $k = \sqrt{\vecb k\cdot \vecb k} =0$ mode.\footnote{I am grateful to Brian Swingle for an illuminating discussion regarding these issues. For important recent progress on theories with multiple Lyapunov exponents, see Ref. \cite{Gharibyan:2018fax}. For a discussion of velocity-dependent Lyapunov exponents, see \cite{Khemani:2018sdn}.} 

At long and intermediate distance scales, $t  \ll \sqrt{\lambda} |\vecb x |$ (recall that $\lambda \gg 1$), the exponential spread of chaos retains the front linear in $|\vecb x|$---the form $e^{\lambda_L(t-|\vecb x|/v_B)}$ with the butterfly velocity $v_B$.  As shown by the perturbative supergravity calculation of the classical shock wave in Section \ref{sec:N4}, this form remains unaffected at least to order $O(1/\lambda^{3/2})$. What we have shown in this paper is that the phenomenon of pole-skipping persists in the inverse perturbative expansion at least to $O(1/\lambda^{3/2})$ and thus, as at infinite coupling, hydrodynamics still displays the same clear analytic signature of the underlying many-body quantum chaos even at finite (large) coupling. A combination of pole-skipping and classical shock wave analyses in $R^2$ and $R^4$ theories shows that, as expected, leading-order $\lambda$-dependent corrections do not affect the maximal Lyapunov exponent. On the other hand, coupling-dependent corrections do shift the value of $k_0$ at the point of chaos $\pc$. Thereby, they also influence the butterfly velocity. Hence, at long distance scales, these observations imply that 
\begin{align}
\lambda_L &= 2\pi T, \\
v_B &= v_B^\infty \left( 1 + \frac{\delta v_B^{(1)}}{\lambda^{(q-1)/2} } + \cdots \right),\label{vBLongDistance}
\end{align}
where $v_B^\infty$ is the butterfly velocity computed at infinite coupling. In this expression, we are thinking of the gravitational $R^q$ corrections as proportional to $1/\lambda^{(q-1)/2}$, so that $\lgb \sim - \alpha' $. A crucial open question at infinite $N_c$ and large but finite $\lambda$ is precisely what types of effects need to be included into the high-energy stringy calculation of \cite{Shenker:2014cwa} in the long-distance (large impact parameter) regime to capture the $\alpha'^3$-dependent effects of classical supergravity that led to Eq. \eqref{vBLongDistance}. 

The discussion of $1/N_c$ corrections is, as is usual in holography, a much more difficult task. This is because beyond the leading order in $1/N_c$ (the order at which conformal central charges fix the $R^2$ coupling), a full analysis requires a computation of quantum gravity loops. Based on Section \ref{sec:GB}, what we can say at present is that in the class of $R^2$ theories, the dominant $1/N_c$ correction does not influence the maximal Lyapunov exponent $\lambda_L = 2 \pi T$, but it reduces the size of the butterfly velocity. 

By combining the results of (perturbative) inverse coupling- and $1/N_c$-dependent observations, we propose the following picture for the relation between hydrodynamics and many-body quantum chaos. At long distance scales, chaos spreads as at $N_c \to \infty$ and $\lambda \to \infty$, i.e. with the exponential (butterfly effect) form $e^{\lambda_L(t-|\vecb x|/v_B)}$. Moreover, the phenomenon of pole-skipping still connects the behaviour of a resummed hydrodynamic mode in the energy density correlator (the sound channel) to the underlying chaotic mixing in precisely the same way as in Refs. \cite{Grozdanov:2017ajz,Blake:2017ris,Blake:2018leo}. The maximal holographic Lyapunov exponent is uncorrected, whereas the $k_0$ of $\pc$ and thereby the butterfly velocity receive both $1/N_c$- and $\lambda$-dependent corrections:
\begin{align}
\lambda_L &= 2\pi T, \\
v_B &= v_B^\infty \left( 1 + \frac{\delta v_B^{(1)}}{\lambda^{(q-1)/2} } - \frac{\delta v_B^{(2)} }{N_c} +  \cdots \right),\label{vBGeneral}
\end{align}
with $\delta v_B^{(n)} > 0$. Beyond $1/N_c$, but for sufficiently large $N_c$, we expect chaos to become characterised by multiple Lyapunov exponents with the front propagation transitioning from ballistic to diffusive spreading. 

This transition was discussed in the recent work by Xu and Swingle \cite{Xu:2018xfz} (see also \cite{Xu:2018dfp,Sahu:2018fuf,Khemani:2018sdn}). There, it was conjectured that, in general, the effects of scrambling spread through the (leading-$\lambda_L$) front that propagates via the following relation (having divided their expression in the exponential by $(\tilde v_B)^{1+p}$): 
\begin{align}
C(t,\vecb x) \sim \exp\left\{- \lambda_L t \left( \frac{1}{ \tilde v_B} \frac{|\vecb x|}{t} - 1 \right)^{1+p}  \right\}.
\end{align}
In the regime of $|\vecb x|  \ll \tilde v_B t$, then,
\begin{align}\label{OTOC-Sw}
C(t,\vecb x) \sim \exp\left\{ \left(1 + i\pi p\right)  \lambda_L \left( t - (1+p) \frac{|\vecb x|}{\tilde v_B}  + \cdots   \right) \right\}.
\end{align}
For $p=0$, this expression reproduces Eq. \eqref{OTOC}. In random unitary circuit models, $p=1$. Now, in light of our above discussion, this phenomenological relation can be fit by using $\tilde v_B = v^\infty_B$ and by incorporating the $1/N_c$-dependent correction to $v_B$, which causes $v_B$ to decrease (cf. Eq. \eqref{vBGeneral}), into $p > 0$. As in Section \ref{sec:N4}, the correlator \eqref{OTOC-Sw} is again complex with a small $1/N_c$-dependent imaginary part. 

As we approach the end of this paper, it is important to remind the reader that, as shown in Section \ref{sec:GB}, pole-skipping and all of its consequences are also valid for all non-perturbative values of the Gauss-Bonnet coupling $\lgb$ in Einstein-Gauss-Bonnet theory. This fact, combined with other results in this paper and those from Refs. \cite{Grozdanov:2017ajz,Blake:2017ris,Blake:2018leo,Haehl:2018izb} strongly reaffirm the statement that there exists a clear and precise analytic connection between hydrodynamics and quantum chaos in large classes of holographic and conformal field theories. Moreover, the widely observed validity of pole-skipping also hints that in all these theories, there exists an underlying symmetry of the type proposed by Ref. \cite{Blake:2017ris}. 

At weak and intermediate coupling, and general $N_c$, the existence of pole-skipping (or its generalisation) is unknown. As noted above, if $N_c$ remains sufficiently large, one may expect diffusive propagation of chaos with multiple Lyapunov exponents. In the absence of small parameters such as $1/N_c$, exponential behaviour is only possible on short time scales \cite{Kukuljan:2017xag}. At weak coupling, when a small parameter can be provided by the (Yang-Mills) coupling constant $g_{YM} = \sqrt{\lambda/N_c}$, the results of \cite{Grozdanov:2018atb} suggest that Lyapunov-type chaos is possible in many-body quantum field theories at late times, again with multiple Lyapunov exponents. Moreover, a close connection between hydrodynamics and many-body quantum chaos has been shown to exist in perturbative quantum field theories with $g_{YM} \ll 1$ for all $N_c$ \cite{Grozdanov:2018atb} (see also Ref. \cite{Stanford:2015owe}). Establishing the precise analytic nature of this connection at the level of precision analogous to the statement of pole-skipping remains an open problem. 

We end by noting that there exists an intriguing connection between the behaviour of $v_B$ and the value of $\eta/s$ in relation to the KSS bound \cite{Kovtun:2003wp,Kovtun:2004de}. As discussed in \cite{Buchel:2008vz}, the qualitative trend of change of $\eta/s$ from the value $1/(4\pi)$ appears to follow a relation, which is analogous to \eqref{vBGeneral}: 
\begin{align}\label{KSS}
\frac{\eta}{s} = \frac{1}{4\pi} \left( 1 + \Delta \right).
\end{align}
For $R^2$ corrections being interpreted as $1/N_c$ corrections, $-1 \ll \Delta < 0 $. Reversing the sign of the $R^2$ coupling (i.e. of $\alpha_3$ multiplying the $R_{\mu\nu\rho\sigma}R^{\mu\nu\rho\sigma}$ term or the Gauss-Bonnet $\lgb$), which reproduces coupling-dependent behaviour, corresponds to having $\Delta > 0$. Consistently, the ``genuine" coupling-dependent corrections to $\CN=4$ SYM theory computed from the $\alpha'^3 R^4$ term also contribute a $\Delta \sim \alpha'^3 \sim 1/\lambda^{3/2} > 0 $.\footnote{See also a review in Ref. \cite{Cremonini:2011iq}. For discussions of higher-derivative corrections to $\eta/s$ at finite chemical potential, see \cite{Myers:2009ij,Cremonini:2009sy,Cremonini:2009ih}. For a recent phenomenological holographic study of coupling-dependent electron flows, see \cite{Erdmenger:2018svl}.} The sign of the change of $\eta/s$ from $1/(4\pi)$ was also shown to be correlated with the trends of changes in positions of higher quasinormal modes (the symmetric branches of non-hydrodynamic poles) in $R^2$ and $R^4$ theories. This was discussed in Ref. \cite{Grozdanov:2016vgg} (see also \cite{Waeber:2015oka}). Thus, it may be tempting to conclude with various speculations regarding the boundedness of quantities (or their ratios) considered in this work, such as an expectation that as a function of the coupling constant, $v_B (\lambda = g_{YM}^2 N_c \to\infty,N_c)  \leq v_B (\lambda, N_c) \leq 1$, with $v_B \to 1$, monotonically, as $g_{YM} \to 0$. Finally, defining a quantity $\mathcal{D} \equiv\CC v_B^2 / \lambda_L$, with $\CC$ independent of $\lambda$, it would then follow that as a function of the coupling constant, $\mathcal{D} \geq \CC v_B^2(\lambda\to\infty) / (2\pi T)$ \cite{Kovtun:2004de,Hartnoll:2014lpa,Maldacena:2015waa,Blake:2016wvh}. The quantity $\mathcal{D}$ shall remain phenomenologically unidentified. 

\acknowledgments{I would like to thank Mike Blake, Richard Davison, Nikos Kaplis, Hong Liu, Mark Mezei, Koenraad Schalm, Vincenzo Scopelliti, Andrei Starinets, Brian Swingle and Phil Szepietowski for illuminating and valuable discussions, and for comments on the draft of this paper. This work was supported by the U. S. Department of Energy under grant Contract Number DE-SC0011090.} 

\bibliographystyle{JHEP}
\bibliography{biblio.bib}
\end{document}